\documentclass[trackchanges]{aastex701}

\newcommand{\htwo}{H$_2$}

\begin{document}

\title{The Goldilocks Molecule: \htwo\ Emission Lines Can Identify Elusive Dwarf AGN}


\author[orcid=0009-0004-2785-4062,gname=Morgan,sname='Micharski']{Morgan Micharski}
\affiliation{Elon University, 100 Campus Drive, Elon, NC 27278, USA}
\email[show]{mmicharski@elon.edu}  

\author[orcid=0000-0002-3703-0719,gname=Chris,sname='Richardson']{Chris Richardson}
\affiliation{Elon University, 100 Campus Drive, Elon, NC 27278, USA}
\email[show]{crichardson17@elon.edu}

\begin{abstract}

We propose using \htwo\ emission lines as a novel diagnostic to identify dwarf AGN by running photoionization models incorporating active intermediate-mass black holes and using an observed WISE dwarf AGN as a template. Though many dwarf AGN would be incorrectly classified as star-forming if the \htwo\ 2.12~$\mu$m/Br$\gamma$ ratio is used, ratios between 
\htwo\ emission lines can reveal active IMBHs when log $n_{\text{H}} \gtrsim 4.0$. This includes the case where $M_{\text{BH}} = 10^3~M_{\odot}$, even though the optical strong lines and NIR coronal lines would likely go undetected. We conclude that \htwo\ emission lines show promise in detecting the most elusive IMBHs.

\end{abstract}

\keywords{\uat{Active galactic nuclei}{16} ---  \uat{Near infrared astronomy}{1093}  ---  \uat{Dwarf galaxies}{416}}


\section{Introduction} 

Intermediate-mass black holes (IMBHs) are a key ingredient in developing a holistic understanding of BH-galaxy co-evolution; however, active IMBHs in dwarf galaxies are relatively elusive, partly due to low accretion rates \citep{Greene2020}, motivating the development of novel methods to locate them. \htwo\ emission has been detected within the interior regions of a sample of massive active galactic nuclei (AGN) \citep{Gillette2026,Costa-Souza2026}. However, \htwo\ emission is weak due to its lack of a permanent dipole moment. Additionally, the metal-poor nature of dwarf AGN inhibits the \htwo\ formation via its main reaction pathway, grain catalysis. Therefore, the potential for \htwo\ to serve as a dwarf AGN diagnostic is poorly understood. Here, we use photoionization models guided by observational constraints to explore whether \htwo\ could serve as an excitation diagnostic for detecting metal-poor dwarf AGN.

\begin{figure*}
\centering
\includegraphics[width=1.0\linewidth]{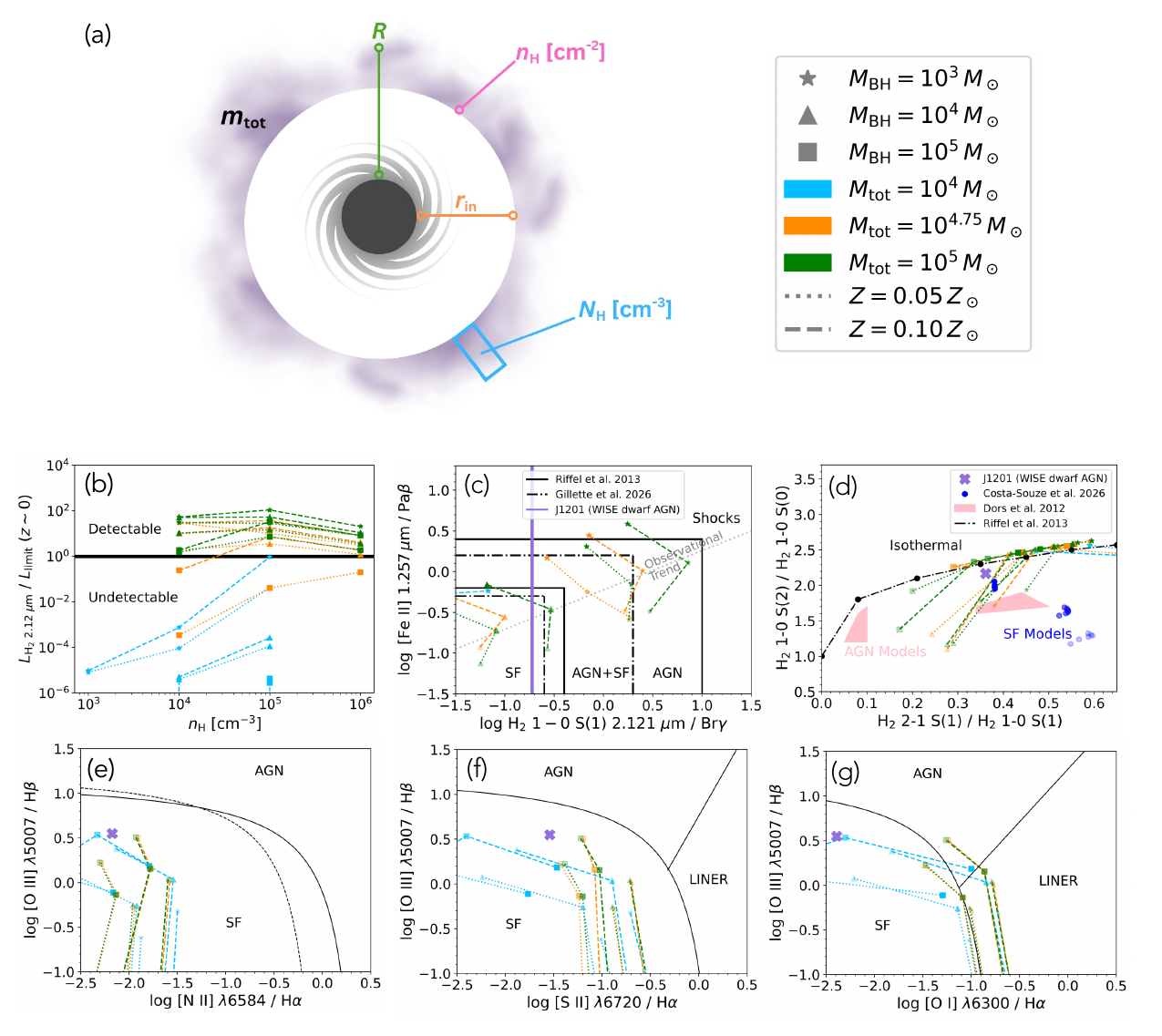}
    \caption{An illustration of the geometric setup of our simulations (a), followed by the detectability of the \htwo~2.12~$\mu$m line with JWST NIRSpec at $z=0.0034$ (b), and several excitation diagnostics in the NIR and optical (c-g). \textit{The diagram in (d) is particularly promising for finding elusive IMBHs.}}
    \label{fig:figure1}
\end{figure*}

\section{Methods} \label{sec:style}

Observations of the dwarf galaxy J120122.30+021108.3 (hereafter, J1201) with the James Webb Space Telescope's Near Infrared Spectrometer (JWST NIRSpec) revealed prominent \htwo\ emission lines \citep{Doan2024}. J1201 is a very low-metallicity ($Z/Z_{\odot}$~=~0.1), nearby ($z = 0.0034$) dwarf identified as a WISE AGN but optically classified as purely star-forming. We use observations of J1201 to set fiducial constraints for our Cloudy 17.02 model \citep{Ferland2017}. We set up a system of equations based on an assumed geometry of the system (Figure 1a):

\begin{equation}
    R = \frac{N_{\text{H}}}{n_{\text{H}}}+r_{\text{in}}
\end{equation}

\begin{equation}
    M_{\text{tot}}\approx\frac{4}{3}\pi\left( R^3-r_{\text{in}}^3 \right)\frac{m_{\text{H}}n_{\text{H}}}{1-Y}
\end{equation}

\noindent We apply the small-angle approximation using the observed redshift and the NIRSpec resolution (0.''1 x 0.''1) to calculate the total radius ($R\approx6.6$~pc) of the dwarf AGN. We select the lower limit of our cloud masses as $m_{tot}=10^{4.0},10^{4.5}, \text{and } 10^{5.0}~M_{\odot}$, by assuming that $m_{tot}$ is approximately set by the mass accreted by the BH throughout cosmic time  \citep{Beckmann2023}. For $m_{tot}=10^{4.75}~\text{and } 10^{5.0}~M_{\odot}$, we vary the hydrogen density at the illuminated face $\log{n_{\text{H}}} = 3.0,~4.0,~5.0$ cm$^{-3}$, while in least massive $m_{tot}$ we explore a lower density range $\log{n_{\text{H}}} = 3.0,~4.0,~5.0$~cm$^{-3}$. We also vary metallicity ($Z/Z_{\odot} = 0.05,~0.1 $) and fix the helium mass fraction $Y=0.25$. We calculate the interior radii ($r_{in}$) and column densities ($N_{H}$) for each input value, with $N_{H}$ as the boundary condition for our simulations. We follow the photoionization model methodology as in \cite{Richardson2025} to set the cloud physical conditions and incident radiation field with the following exceptions: (1) we enable the full set of \htwo\ and Fe~II energy levels; (2) instead of setting an ionization parameter, the bolometric luminosity ($L_{\text{bol}}$) is set by assuming an accretion rate ($\dot{m}$) of $0.1~\dot{m}_{\text{edd}}$ and radiative efficiency $\epsilon_{\text{rad}} = 0.1$ for black hole masses of $M_{\text{BH}}$ = $10^{3}$, $10^{4}$, \text{and} $10^{5}$ $M_{\odot}$.

\section{Results} \label{sec:floats}

Figure~1b shows that the simulations with $m_{tot}=10^{4}~M_{\odot}$ generally predict \htwo\ 2.12~$\mu$m luminosities several orders of magnitude lower than the NIRSpec sensitivity limit ($\sim2 \times 10^{-18}$~erg~s$^{-1}$~cm$^{-2}$). In contrast, our simulations with $m_{tot}=10^{5}~ M_{\odot}$  predict higher \htwo\ 2.12~$\mu$m luminosities than the NIRspec limiting luminosity, while the $m_{tot}=10^{4.75}~ M_{\odot}$ simulations produce a mixture of detectable and undetectable emission. Despite the lower dust-to-gas ratios in dwarfs, we find that \htwo\ still forms through grain catalysis rather than associate detachment for simulations with $m_{tot}=10^{4.75}$,~$10^{5}~M_{\odot}$. 

We analyzed the resulting \htwo\ spectra using two NIR diagnostic diagrams that can distinguish star-forming regions from massive AGN. Figure~1c uses $\text{H}_{2}~2.121 ~\mu \text{m}/\text{Br} \gamma$ vs. $[\text{Fe II}]~1.257~\mu \text{m}/\text{Pa} \beta$ to distinguish AGN, star-forming, and composite sources \citep{Riffel2013,Gillette2026}. According to the $\text{H}_{2}~2.121 ~\mu \text{m}/\text{Br} \gamma$ ratio alone, J1201 registers as a star-forming galaxy unless [Fe~II]/Pa$\beta$ is highly elevated relative to most observations. We find that our simulations with $m_{tot}=10^{4} ~M_{\odot}$ do not classify as AGN, on account of the cloud not having enough mass to yield significant \htwo\ formation and emission (Figure 1b). Additionally, only simulations with $m_{tot}=10^{4.75},~10^{5}M_{\odot}$ and $M_{\text{BH}}=10^{3} M_{\odot}$  were routinely classified as either composite sources or AGN for both metallicities, respectively. Interestingly, the simulations with these cloud masses at $0.05~Z_{\odot}$ with $M_{\text{BH}}=10^{3} M_{\odot}$ and at $0.1~Z_{\odot}$ with $M_{\text{BH}}=10^{4} M_{\odot}$ fall in the star-forming region.

Figure 1d shows a diagram of two \htwo\ emission line ratios, with J1201 classified as an AGN because its \htwo\ emission lies just below the isothermal gas distribution line, indicative of AGN activity \citep{Riffel2013,Costa-Souza2026}. Our only simulation characterized as an AGN with $m_{tot}=10^{4} M_{\odot}$ requires the lowest $M_{\text{BH}}$ and $Z$ values we consider but the highest $n_{\text{H}}$. \textit{However, the only simulations with $m_{tot}=10^{4.75},~10^{5.0}~ M_{\odot}$ not characterized as an AGN have lowest hydrogen densities ($n_{H}=10^{4}~\text{cm}^{-3}$)}. 

Figures~1e-g investigate three optical diagnostic diagrams that together can recover AGN in dwarfs with strong emission lines \citep{Polimera2022}. Our models are classified as star-forming except in the [O~I]/H$\alpha$ plot. The simulations with $m_{tot}=10^{4.75}, 10^{5.0}~M_{\odot}$ and $M_{\text{BH}}=10^{3} M_{\odot}$ do not appear on these diagrams because their weak [O~III] emission, which we determined would be undetectable with SDSS at $z=0.0034$.

\section{Discussion}

\begin{itemize}
    \item The diagram in Figure~1c suggests that many low-metallicity dwarf AGN could be incorrectly classified as star-forming.
    \item Figure~1d indicates that $\log{n_{\text{H}}} \gtrsim 4.0$ is required to yield \htwo\ emission line ratios that separate dwarf AGN and star-forming models; however, the AGN region is also occupied by low-density shock models \citep{Costa-Souza2026}
    \item The low [O~III]/H$\beta$ from our simulations, relative to J1201 (Figures~1e-g), suggest that either a higher $M_{\text{BH}}$ or $\dot{m}$ should be explored to increase $L_{\text{bol}}$. Alternatively, stellar radiation could easily generate the required [O~III]/H$\beta$, while the \htwo\ spectrum is reproduced by an active IMBH.
    \item Intriguingly, simulations with $M_{\text{BH}}=10^{3}~M_{\odot}$ have [O~III]~$\lambda$5007 and coronal line luminosities too low to be detected with SDSS and NIRSpec, respectively; yet, the \htwo\ spectrum is detectable and contains active IMBH signatures. In this situation, the optical spectrum of a dwarf AGN could be due to starlight while the AGN dominates \htwo\ excitation. This suggests that \htwo\ emission line ratios are a potentially promising avenue for identifying elusive IMBHs in the lower mass gap \citep{Burke2026}.
\end{itemize}

\begin{acknowledgements}
    We thank Vianney Lebouteiller and Nick Abel for helpful comments. MM acknowledges the North Carolina's GlaxoSmithKline Women in Science Scholars Program. CR acknowledges NASA grant 24-RIA24-0038. 
\end{acknowledgements}

\bibliography{sample701}{}
\bibliographystyle{aasjournalv7}



\end{document}